\documentclass[12pt, a4paper] {article}
\usepackage{amsfonts}
\usepackage{verbatim}
\usepackage{latexsym}
\usepackage{amsmath}

\begin{document}

\begin{flushright} arXiv: 1003.3390 [hep-th]\\ CAS-PHYS-BHU Preprint
\end{flushright} 

\vskip 0.1cm

\begin{center}
{\sf \large Absolutely anticommuting (anti-)BRST symmetry transformations
for topologically massive Abelian gauge theory}\\

\vskip 2.5cm

{\sf S. Gupta$^{(a)}$, R. Kumar$^{(a)}$, R. P. Malik$^{(a, b)}$}\\
{\it $^{(a)}$Physics Department, Centre of Advanced Studies,}\\
{\it Banaras Hindu University, Varanasi - 221 005, (U.P.), India}\\

\vskip 0.1cm

{\bf and}\\

\vskip 0.1cm

{\it $^{(b)}$DST Centre for Interdisciplinary Mathematical Sciences,}\\
{\it Faculty of Science, Banaras Hindu University, Varanasi - 221 005, India}\\
{\small {e-mails: guptasaurabh4u@gmail.com, raviphynuc@gmail.com, malik@bhu.ac.in}}
\end{center}

\vskip 1.5cm

\noindent
{\sf Abstract:} We demonstrate the existence of the nilpotent and absolutely anticommuting 
Becchi-Rouet-Stora-Tyutin (BRST) and anti-BRST symmetry transformations for the four 
(3 + 1)-dimensional (4D) topologically massive Abelian $U(1)$ gauge theory 
that is described by the coupled Lagrangian densities (which incorporate
the celebrated $ (B \wedge F) $ term). The absolute anticommutativity of 
the (anti-) BRST symmetry transformations is ensured by the existence of 
a Curci-Ferrari type restriction that emerges from the 
superfield formalism as well as from the equations of motion that are derived from the 
above coupled Lagrangian densities. We show the invariance of the action from the point 
of view of the symmetry considerations as well as superfield formulation. We discuss, furthermore,
the topological term within the framework of superfield formalism and
provide the geometrical meaning of its invariance under the
(anti-) BRST symmetry transformations.

\vskip 1 cm



\newpage

\noindent
{\bf 1. Introduction}\\

\noindent
A couple of decisive mathematical 
features, that are closely connected with the basic concepts of 
Becchi-Rouet-Stora-Tyutin (BRST) formalism [1-4], are

(i) the nilpotency $(s_{(a)b}^2 = 0, Q_{(a)b}^2 = 0) $ of the (anti-) BRST symmetry 
transformations $(s_{(a)b})$ and their corresponding generators ($Q_{(a)b}$), and

(ii) the absolute anticommutativity $(s_b s_{ab} + s_{ab} s_b = 0,\; 
Q_b Q_{ab} + Q_{ab} Q_b = 0 )$ of the (anti-) BRST symmetry transformations 
$s_{(a)b}$ (in their operator form) and the generators $Q_{(a)b}$ which
generate the transformations $s_{(a)b}$.

The former mathematical 
property physically implies the fermionic nature of $s_{(a)b}$ (as well as $Q_{(a)b}$)
and the latter property encodes the linear independence of $s_b$ {\it vis-{\`a}-vis}
$s_{ab}$ (and $Q_b$ versus $Q_{ab}$). These mathematical properties are very sacrosanct 
and they must be obeyed in the BRST description of any arbitrary gauge/reparametrization 
invariant theories.

The role of the BRST formalism is quite significant in the description of the non-Abelian
1-form gauge theories which are at the heart of theoretical foundations of the standard
model of high energy physics. Despite stunning success stories associated with the standard
model, its shortcomings are the detection of the mass of the neutrino and 
no experimental evidence for the Higgs boson (so far!). One of the roles of the Higgs particle is to
generate suitable masses for the gauge particles and fermions. Thus, its detection is 
very crucial for the sanctity of the theoretical foundations of the standard model. Since the
esoteric Higgs bosons have not yet been seen experimentally, some alternative models have been
proposed for the mass generation, symmetry breaking, etc. One of the
alternate models, for the mass generation of the gauge fields, is the inclusion
of the topological ($B \wedge F$) term in the Lagrangian density of the 1-form and 2-form
(non-) Abelian gauge theories where the mass generation of the 1-form gauge
boson is very natural [5-8].

The 2-form [$B^{(2)} = (1/2!) (dx^\mu \wedge dx^\nu) B_{\mu\nu}$] antisymmetric tensor gauge field
$B_{\mu\nu}$ [9,10] has become quite popular because of its relevance in the 
context of superstring [11,12] and supregravity theories [13]. Besides being a theoretical
generalization of the 1-form gauge field [14], it provides the field theoretic models for 
the Hodge theory [15-17] and it is also relevant in the context of condensed matter 
physics [18]. Its constraint structures [19], BRST quantization scheme [20-22], etc.,
have been studied. These studies have led to
some novel features (that are found to be {\it absent} in the study
of 1-form (non-) Abelian gauge theories). Thus, the (non-) Abelian 2-form gauge field
is endowed with a very rich mathematical and theoretical structures.

In our present endeavor, we focus on the gauge theory
of the Abelian 1-form and 2-form gauge fields
that are coupled with each-other through the famous $(B \wedge F)$ term.
In fact, in the presence of the $(B \wedge F)$ term, we study the present (4D
topological massive Abelian $U(1)$ gauge) model within the framework of BRST formalism.
We promote the gauge symmetry of the theory to the off-shell nilpotent
and absolutely anticommuting (anti-) BRST symmetry transformations where the basic tenets
of the BRST formalism are fully respected. We find some novel features in our study.
These are

(i) the existence of the coupled Lagrangian densities for the description of an Abelian
gauge theory that incorporates the Abelian 1-form and 2-form gauge fields 
together with the topological term. This observation is novel in the sense that it is very
similar to the case of the non-Abelian 1-form gauge theory where such a kind of Lagrangian
densities do exist [23,24],

(ii) the derivation of the Curci-Ferrari (CF) type restriction from the coupled Lagrangian
densities as well as from the superfield approach to BRST formalism. This aspect
of our present Abelian theory is exactly same as the one observed in the case of
non-Abelian 1-form gauge theory [23,24] (where, for the first time, CF condition 
appeared [25]), and

(iii) the interpretation of the topological $(B \wedge F)$ term within the framework
of the superfield approach to BRST formalism and its geometrical meaning {\it vis-{\`a}-vis}
the rest of the terms of the theory.

The key factors that have contributed to our main motivation for present
investigation are as follows.  
First and foremost, the BRST construction, for our present model,
has been found to be endowed with the BRST symmetries that are non-nilpotent 
(see, e.g. [26]). Thus, it is an interesting endeavor for us to obtain the symmetries 
that obey the key requirements of the BRST
formalism. Second, we demonstrate that our present model is described by a coupled set
of Lagrangian densities due to the existence of CF-type restriction. Third, it is important for us
to check the relevance of our earlier work [27] in the context of our present model
which is more general than the BRST description of the {\it free} Abelian 2-form gauge theory.
Fourth, our present 4D theory provides a field theoretic model where the superfield and Lagrangian
approaches to the BRST formalism blend together in a useful and clarifying manner.
Finally, the {\it non-Abelian} generalization of the present model has been a topic of 
intense research for quite sometime [8,28,29]. We wish to generalize our present model to 
the non-Abelian case by exploiting the superfield formalism proposed by Bonora, etal. 
[30,31].

Our present paper is organized as follows. In Sec. 2, we discuss about the gauge 
symmetries and constraint structures of the 4D massive Abelian $U(1)$ gauge theory. 
Our Sec. 3 is devoted to the discussion of 
the on-shell nilpotent BRST and anti-BRST symmetry transformations for a {\it single}
Lagrangian density and we demonstrate that these transformations are non-anticommuting
in nature. In Sec. 4, we provide a brief synopsis of the superfield approach 
to derive the proper and precise (anti-) BRST
symmetry transformations. In Secs. 5 and 6, we dwell a bit on the (anti-) BRST invariance 
of the present theory within the frameworks of the Lagrangian and superfield 
formalisms, respectively. Finally, we make some concluding remarks 
and point out a few future directions in Sec. 7. \\

\noindent
{\bf 2. Preliminaries: gauge symmetries and constraints}\\

\noindent
We begin with the Lagrangian density of a massive gauge invariant Abelian model in four
(3 + 1)-dimensions of spacetime. This Lagrangian density 
incorporates the celebrated topological ($ B \wedge F $) term as given below\footnote {We 
adopt here the flat metric ($\eta_{\mu\nu}$) with signatures
($ +1, -1, -1, -1$) so that $ P \cdot Q = \eta_{\mu\nu} P^\mu Q^\nu
= P_0 Q_0 - P_i Q_i $ is the dot product between two non-null vectors
$ P_\mu $ and $Q_\mu$ where $\mu, \nu, \eta, \kappa, . . . . = 0, 1, 2, 3$
correspond to the spacetime directions and $ i, j, k, . . . . = 1, 2, 3$
stand for the space directions only. We make the choice $\varepsilon_{0123}
= +1 = - \varepsilon^{0123}$ for the totally antisymmetric Levi-Civita
tensor ($\varepsilon_{\mu\nu\eta\kappa}$) that obeys $\varepsilon_{\mu\nu\eta\kappa} \varepsilon^{\mu\nu\eta\kappa} 
= - 4 !$, $\varepsilon_{\mu\nu\eta\kappa}  \varepsilon^{\mu\nu\eta\sigma} = - 3 ! \;
\delta_\kappa^\sigma$, etc. The component $\varepsilon_{0ijk}
= \epsilon_{ijk}$ is the 3D Levi-Civita tensor.} 
\begin{eqnarray}
{\cal L}_0 = - \; \frac {1} {4} \; F^{\mu\nu} \; F_{\mu\nu} + \frac {1}{12} \; H^{\mu\nu\eta}
\; H_{\mu\nu\eta} + \frac {m}{4} \; \varepsilon^{\mu\nu\eta\kappa} \; F_{\mu\nu} 
\; B_{\eta\kappa},
\end{eqnarray}
where the totally antisymmetric quantities $F_{\mu \nu} = \partial_\mu A_\nu 
- \partial_\nu A_\mu$ and 
$H_{\mu \nu \eta} = \partial_\mu B_{\nu \eta} + \partial_\nu B_{\eta \mu}
+ \partial_\eta B_{\mu \nu}$ are the curvature tensors owing their origin to 2-form
$F^{(2)} = d A^{(1)} \equiv \frac {1}{2!} \; (d x^\mu \wedge d x^\nu) F_{\mu \nu}$ and 3-form 
$H^{(3)} = d B^{(2)} \equiv \frac {1}{3!}\; (d x^\mu \wedge d x^\nu \wedge d x^\eta) 
H_{\mu \nu \eta}$, respectively. Here $d = d x^\mu \; \partial_\mu$  (with $d^2 = 0$) is the 
exterior derivative and 1-form $A^{(1)} = d x^\mu A_\mu$ and 2-form $B^{(2)} = \frac {1}{2!} 
\;(d x^\mu \wedge d x^\nu)\;  B_{\mu \nu}$ define the Abelian 4-vector $(A_\mu)$ and second 
rank anti-symmetric $(B_{\mu \nu} = - B_{\nu \mu})$ tensor ($B_{\mu \nu}$) gauge fields. 
It is clear that the parameter `$m$' has the dimension of mass in the physical four
dimensions of spacetime.

The above Lagrangian density transforms to a total spacetime derivative 
(i.e. $ \delta_{(gt)}\; {\cal L}_0 = \partial_\mu [ m \; \varepsilon^{\mu \nu \eta \kappa}
\Lambda_\nu \; (\partial_\eta A_\kappa) ]$) under the following infinitesimal gauge 
transformations\footnote{Note that the Lagrangian density ${\cal L}_0$ 
respects a couple of independent gauge symmetry transformations: 
(i) $A_\mu \to A^\prime_\mu = A_\mu, \; B_{\mu\nu} \to B_{\mu\nu}^\prime 
= B_{\mu\nu} + (\partial_\mu \Lambda_\nu 
- \partial_\nu \Lambda_\mu)$, and (ii) $A_\mu \to A^\prime_\mu
= A_\mu + \partial_\mu \Lambda, \; B_{\mu\nu} \to B_{\mu\nu}^\prime = B_{\mu\nu}$.
For the sake of generality, however, we have taken the combination of these two 
transformations {\it together} in (2).} $\delta_{(gt)}$ (with gauge parameter $\Lambda$ and 
$\Lambda_\mu$)
\begin{eqnarray}
\delta_{(gt)} A_\mu = \partial_\mu \Lambda, \hskip 2cm \delta_{(gt)} B_{\mu \nu} 
= \partial_\mu \Lambda_\nu - \partial_\nu \Lambda_\mu.
\end{eqnarray} 
Thus, the action $S_0 = \int d^4 x \; {\cal L}_0$ remains invariant under 
the infinitesimal gauge transformations (2). It is straightforward to check 
that the following Euler-Lagrange equations of motion
\begin{eqnarray}
\partial_\mu F^{\mu \nu} = \frac {1}{2} \; m \; \varepsilon^{\mu \nu \eta \kappa}
\partial_\mu B_{\eta \kappa}, \qquad
\partial_\mu H^{\mu \nu \eta} = \frac {1}{2} \; m \; 
\varepsilon^{\nu \eta \kappa \sigma} F_{\kappa \sigma},
\end{eqnarray}
emerge from the Lagrangian density (1). The components of the conjugate momenta 
with respect to the vector field $A_\mu$ and tensor field $B_{\mu \nu}$: 
\begin{eqnarray}
&&\Pi^{0}_{(A)} = 0,  \hskip 2cm \Pi^{i}_{(A)} = - F^{0 i} + \frac {1}{2} \; m \;
\varepsilon^{0 i j k} B_{j k},\nonumber\\
&&\Pi^{0 i}_{(B)} = 0, \hskip 2cm \Pi^{i j}_{(B)} = \frac {1}{2} H^{0 i j}, 
\end{eqnarray}
ensure that $\Pi^{0}_{(A)} \approx 0, \; \Pi^{0 i}_{(B)} \approx 0$ are the 
primary constraints on the theory. As a consequence, the equations of motion 
with respect to $A_0$ field and $B_{0 i}$ field (see, e.g. [32] for details):
\begin{eqnarray}
\partial_i \Bigl( F_{0 i} - \frac {1}{2} \; m \; \epsilon_{i j k} B_{j k} \Bigr)
&\equiv & - \; \partial_i \Pi^{(A)}_i  \approx 0, \nonumber\\ 
\partial_j H_{0 i j} + \frac {1}{2} \; m \; \epsilon_{i j k} F_{j k}  
&\equiv & 2 \; \partial_j \Pi^{(B)}_{i j} + \frac {1}{2} \; m \; 
\epsilon_{i j k} F_{j k} \approx 0, 
\end{eqnarray}
lead to the derivation of the secondary constraints on the theory. The above primary
and secondary constraints are the first-class constraints in the language of Dirac's 
prescription for the classification scheme [33,34].

The continuous gauge symmetry transformations (2) lead to the derivation of the  
Noether conserved current as given below:
\begin{eqnarray}
J^\mu_{(gt)} = \frac {1}{2} \; m \; \varepsilon^{\mu \nu \eta \kappa} \;(\partial_\nu \Lambda) 
B_{\eta \kappa} - F^{\mu \nu} \; \partial_\nu \Lambda + H^{\mu \nu \eta} \; 
\partial_\nu \Lambda_\eta - m \; \varepsilon^{\mu \nu \eta \kappa} \Lambda_\nu 
\; (\partial_\eta A_\kappa),
\end{eqnarray}
because $\partial_\mu J^\mu_{(gt)} = 0$ when we exploit the 
Euler-Lagrange equations of motion (3). 
The conserved charge (i.e. $Q_{(gt)} = \int d^3 x \; J^0_{(gt)}$)
\begin{eqnarray}
Q_{(gt)} &=& \int d^3 x \Bigl[ \Bigl ( F_{0 i} - \frac {1}{2} \; m \; \epsilon_{i j k} 
B_{j k} \Bigr ) \partial_i \Lambda + 2 \; \Pi^{i j}_{(B)} (\partial_i \Lambda_j ) 
+ m \; \epsilon_{i j k} \; \Lambda_i \; (\partial_j A_k) \Bigr ]\nonumber\\
&\equiv & \int d^3 x \Bigl[ \Pi^i_{(A)} (\partial_i \Lambda ) + 2 \; \Pi^{i j}_{(B)} 
(\partial_i \Lambda_j) + m \; \epsilon_{i j k} \; \Lambda_i \; (\partial_j A_k) \Bigr ],
\end{eqnarray}
generates the following transformations with the help of (11) (see below)
\begin{eqnarray}
\delta_{(gt)} A_i &=& - i \; [A_i, \; Q_{(gt)}]  \; = \; \partial_i \Lambda, \nonumber\\
\delta_{(gt)} B_{i j} &=& - i \; [B_{i j}, \; Q_{(gt)}]  \; = \; \partial_i \Lambda_j - 
\partial_j \Lambda_i.
\end{eqnarray}
Thus, the Noether conserved charge $Q_{(gt)}$ does {\it not} generate all the 
transformations for {\it all} the components of the field. For instance, we can never be able to
obtain the transformations for the components $A_0$ and $B_{0i}$ of the 1-form
and 2-form gauge fields, respectively, from the above charge $Q_{(gt)}$.

The basic tenet of gauge theory ensures that all the gauge transformations
should be generated by the first-class constraints of the theory [35]. Such,
a generator $(G)$, in terms of the above first-class constraints, 
is\footnote{It will be noted that one of the secondary constraints (cf. (5)) includes
the topological term ``$m \epsilon_{ijk} F_{jk}$'' as well. However, this term does not generate
any transformation. Thus, we have not incorporated this term in the expression for $G$ so
that we could get a compact and simple form of $G$. In principle, this term should
be present in our expression for $G$.}
\begin{eqnarray}
G &=& \int d^3 x \; \Bigl [ (\partial_0 \Lambda) \; \Pi^0_{(A)} + \Lambda \; \partial_i \Pi^i_{(A)}
+ (\partial_0 \Lambda_i - \partial_i \Lambda_0 ) \; \Pi^{0i}_{(B)} \nonumber\\ 
&+& (\partial_i \Lambda_j - \partial_j \Lambda_i ) \; \Pi^{i j}_{(B)} \Bigr ].
\end{eqnarray}
The above generator leads to the derivation of (2) if we exploit 
the following general rule for the transformation of the generic field $\Phi$, namely;
\begin{eqnarray}
\delta_{(gt)} \Phi \; = \; - i \; \bigl[\Phi, \; G \bigr], \qquad \Phi = A_\mu, B_{\mu \nu},
\end{eqnarray}
supplemented with the following canonical commutation relations
\begin{eqnarray}
&& \bigl[A_0 ({\bf x}, t), \; \Pi^0_{(A)} ({\bf y}, t)\bigr] \; = \; i \; \delta^{(3)} 
({\bf x - y}), \nonumber\\
&& \bigl[A_i ({\bf x}, t), \; \Pi^j_{(A)} ({\bf y}, t)\bigr] \; = \; i \; \delta^j_i \;
\delta^{(3)} ({\bf x - y}),\nonumber\\
&& \bigl[B_{0 i} ({\bf x}, t), \; \Pi^{0 j}_{(B)} ({\bf y}, t)\bigr] \; = \; i \; \delta^j_i \;
\delta^{(3)} ({\bf x - y}),\nonumber\\
&& \bigl[B_{i j} ({\bf x}, t), \; \Pi^{k l}_{(B)} ({\bf y}, t)\bigr] \; = \; \frac {i}{2} \; 
(\delta^k_i \; \delta^l_j - \delta^l_i \; \delta^k_j)\; \delta^{(3)} ({\bf x - y}),  
\end{eqnarray}
and all the rest of the brackets should be taken to be zero.

At this stage, a couple of key points are to be noted. First, neither the conserved 
charge $Q_{(gt)}$ nor the generator $G$ produces the residual symmetry 
transformation\footnote {In other words, if we assume that the gauge 
parameter $\Lambda_\mu$ is a field that transforms as $\delta_\omega \;\Lambda_\mu 
= \partial_\mu \omega$ under a residual gauge transformation $\delta_\omega$, then also, 
the Lagrangian density remains invariant. We shall see later that the parameter
$\Lambda_\mu$ would be identified with the (anti-) ghost fields $(\bar C_\mu) C_\mu$ 
within the framework of BRST formalism (see, Sec. 3).} 
that is present in the gauge transformations $\delta_{(gt)} B_{\mu \nu} 
= \partial_\mu \Lambda_\nu - \partial_\nu \Lambda_\mu$ when $\Lambda_\mu \to \Lambda_\mu 
+ \partial_\mu \omega$. 
Second, according to the Dirac's prescription for the quantization of system with 
constraints, we must demand that the physical states of the theory should be 
annihilated by the first-class constraints (and the ensuing conditions should remain 
invariant with respect to the time evolution of the system). We do not obtain 
these conditions from $Q_{(gt)}$ and $G$ (unless we impose the same, by 
hand, from outside).

The resolutions of these important issues could be addressed
within the framework of BRST formalism. This is what precisely 
we envisage to do in our forthcoming sections. We also comment on various 
subtle issues that are associated with the BRST and superfield 
formulation of the topologically massive Abelian model which is
under consideration in our present endeavor.\\

\noindent
{\bf 3. On-shell nilpotent (anti-) BRST invariant Lagrangian density 

and comments on the covariant canonical quantization}\\

\noindent
To answer the above raised issues, we begin with a generalized version of 
Lagrangian density ${\cal L}_0$ which incorporates the gauge-fixing terms 
(in the Feynman gauge) and Faddeev-Popov ghost terms as [26]
\begin{eqnarray}
{\cal L}_b &=& - \; \frac {1}{4} F^{\mu \nu} F_{\mu \nu} + \frac {1}{12} H^{\mu \nu \eta}
H_{\mu \nu \eta} + \frac {1}{4} \; m \; \varepsilon^{\mu \nu \eta \kappa}
B_{\mu \nu} F_{\eta \kappa} - \frac {1}{2} (\partial \cdot A)^2 \nonumber\\
&-& \frac {1}{2} (\partial^\nu B_{\nu \mu} - \partial_\mu \phi )^2 - \frac {1}{2}
(\partial \cdot \bar C)(\partial \cdot C)
- i \; \partial_\mu \bar C \; \partial^\mu C \nonumber\\
&-& (\partial_\mu \bar {C_\nu} - \partial_\nu \bar {C_\mu}) \; (\partial^\mu C^\nu)
+ \partial_\mu \bar \beta \; \partial^\mu \beta, 
\end{eqnarray} 
where the fermionic (anti-) ghost fields $(\bar C_\mu) C_\mu$ with 
$(\bar C_\mu^2 = C_\mu^2 = 0, C_\mu \bar C_\nu + \bar C_\nu C_\mu = 0, C_\mu C_\nu +
C_\nu C_\mu = 0,$ etc.) are the generalization of the gauge parameter $\Lambda_\mu$ and
the bosonic (anti-) ghost fields $(\bar \beta) \beta$ are the generalization of the 
gauge parameter `$\omega$' (that was present in the symmetry transformation 
$\Lambda_\mu \to \Lambda_\mu + \partial_\mu \omega$). In exactly similar fashion, 
the gauge parameter $\Lambda$ has been replaced by the fermionic $(C^2 = \bar C^2 = 0, 
C \bar C + \Bar C C = 0)$ (anti-) ghost fields $(\bar C) C$. It is self-evident that 
$(\bar C_\mu ) C_\mu$ and $(\bar C) C$ have ghost number equal to $(- 1)+1$
and $(\bar \beta) \beta$ have ghost number $(- 2)+2$, respectively.

The gauge-fixing term $(\partial^\nu B_{\nu \mu})$ for the 2-form gauge field has its
origin in the co-exterior derivative $\delta = - * \; d \; *$ where $*$ is the Hodge duality
operation on the 4D spacetime manifold. It can be readily checked that: 
$\delta B^{(2)} = - * \;d * B^{(2)} = (\partial^\nu B_{\nu \mu}) \; dx^\mu$ (i.e. a 
1-form). There is a room, however, for adding/subtracting a 1-form to this. This can be 
constructed with a massless $(\Box \phi = 0)$ scalar field $(\phi)$ by exploiting an 
exterior derivative (i.e. $F^{(1)} = dx^\mu \; \partial_\mu \phi$) (see, e.g. [17] 
for details). This has been done in the above with a minus sign for algebraic convenience.
It serves the purpose of stage-one reducibility in the theory (which was {\it not} incorporated
in\footnote{It is precisely because of this reason that the BRST transformations, quoted 
in [26], are {\it not} on-shell nilpotent of order two.} [26]). The gauge-fixing 
terms for 1-form (anti-) ghost fields $(\bar C_\mu ) C_\mu $
as well as Abelian $U (1)$ gauge field $A_\mu$ have been taken into account in 
${\cal L}_b$ by incorporating $(- \frac{1}{2} \; \bigl(\partial \cdot \bar C)(\partial \cdot C)
\bigr)$ and  $(- \frac {1}{2} \; \bigl(\partial \cdot A)^2 \bigr)$ terms, respectively. 
These (with ghost number zero), too, owe their origin to the co-exterior derivative 
$\delta = - * d *$.

The above Lagrangian density ${\cal L}_b$ respects the following nilpotent 
$\tilde s_{(a)b}^2 = 0$ (anti-) BRST symmetries $(\tilde s_{(a)b})$ on the on-shell 
$ \bigl(\Box \beta = 0, \; \Box \bar \beta = 0, \; \Box C_\mu 
- \frac {1}{2} \partial_\mu (\partial \cdot C) = 0, \; \Box \bar C_\mu 
- \frac {1}{2} \partial_\mu (\partial \cdot \bar C) = 0, \; \Box \phi = 0 \bigr)$. The explicit 
form of these transformations (as operators on the fields) are 
\begin{eqnarray}
&&\tilde s_b A_\mu = \partial_\mu C, \quad  \tilde s_b \bar C = - \; i \; (\partial \cdot A),
\quad \tilde s_b B_{\mu \nu} = (\partial_\mu C_\nu - \partial_\nu C_\mu), \nonumber\\
&& \tilde s_b C_\mu = \partial_\mu \beta, \quad \tilde s_b \bar C_\mu 
= (\partial^\nu B_{\nu \mu} - \partial_\mu \phi ), 
\quad \tilde s_b \phi = - \; \frac {1}{2}\; (\partial \cdot C), \nonumber\\
&& \tilde s_b \bar \beta = + \; \frac {1}{2}\; (\partial \cdot \bar C), \quad 
\tilde s_b C = 0, \quad \tilde s_b \beta = 0,
\end{eqnarray}
\begin{eqnarray}
&&\tilde s_{ab} A_\mu = \partial_\mu \bar C, \quad  \tilde s_{ab} C 
= + \; i \; (\partial \cdot A),
\quad \tilde s_{ab} B_{\mu \nu} = (\partial_\mu \bar C_\nu 
- \partial_\nu \bar C_\mu), \nonumber\\
&& \tilde s_{ab} \bar C_\mu = \partial_\mu \bar \beta, \quad 
\tilde s_{ab} C_\mu = - \; (\partial^\nu B_{\nu \mu} - \partial_\mu \phi ), 
\quad \tilde s_{ab} \phi = - \; \frac {1}{2}\; (\partial \cdot \bar C), \nonumber\\
&& \tilde s_{ab} \beta = - \; \frac {1}{2}\; (\partial \cdot C), \quad 
\tilde s_{ab} \bar C = 0, \quad  \tilde s_{ab} \bar \beta = 0.
\end{eqnarray}
In fact, it can be checked that the Lagrangian density ${\cal L}_b$ transforms 
(to the total spacetime derivatives) under the above transformations as 
\begin{eqnarray}
\tilde s_b {\cal L}_b &=& - \; \partial_\mu \Bigl[(\partial^\mu C^\nu - \partial^\nu C^\mu)
(\partial^\sigma B_{\sigma \nu} - \partial_\nu \phi ) 
+ \; \frac {1}{2} \; (\partial_\nu B^{\nu \mu} 
- \partial^\mu \phi)(\partial \cdot C)\nonumber\\
&+& (\partial \cdot A) \; \partial^\mu C - \; \frac {1}{2} \; 
(\partial \cdot \bar C) \; \partial^\mu \beta  
- m \; \varepsilon^{\mu \nu \eta \kappa} \; C_\kappa (\partial_\eta A_\kappa) \Bigr], 
\end{eqnarray}  
\begin{eqnarray}
\tilde s_{ab} {\cal L}_{b} &=& - \; \partial_\mu \Bigl[(\partial^\mu \bar C^\nu 
- \partial^\nu \bar C^\mu) (\partial^\sigma B_{\sigma \nu} - \partial_\nu \phi ) 
+ \; \frac {1}{2} \; (\partial_\nu B^{\nu \mu} 
- \partial^\mu \phi)(\partial \cdot \bar C)\nonumber\\
&+& (\partial \cdot A) \; \partial^\mu \bar C + \; \frac {1}{2} \; 
(\partial \cdot C) \; \partial^\mu \bar \beta  
- m \; \varepsilon^{\mu \nu \eta \kappa} \; \bar C_\kappa (\partial_\eta A_\kappa) \Bigr]. 
\end{eqnarray}
As a consequence, the  action $S = \int d^4 x \; {\cal L}_b$ remains invariant 
under the nilpotent symmetry transformations $\tilde s_{(a)b}$.

We close this section with the following remarks. First, using the Noether's theorem, 
one can compute the (anti-) BRST charges $\tilde Q_{(a)b}$ which turn out to be 
conserved and nilpotent. Second, the physicality criteria $\tilde Q_{(a)b} |\; phys > = 0$
lead to the annihilation of the physical states $|\; phys >$  by the operator 
form of the first-class constraints (4) and (5).
Third, the analogue of the  gauge transformations (2) and residual gauge transformations 
$(\delta_\omega \Lambda_\mu = \partial_\mu \omega \Rightarrow \tilde s_b C_\mu 
= \partial_\mu \beta, \; \tilde s_{ab} \bar C_\mu = \partial_\mu \bar \beta)$ are 
generated by the nilpotent and conserved charges $\tilde Q_{(a)b}$. Fourth, it 
can be checked that each basic field of the theory has its corresponding 
canonical momentum. As a consequence, one can perform the {\it covariant} canonical quantization
of the theory in a straightforward manner. Finally, despite the 
above cited good features, it can be checked that the above symmetry transformations 
do not satisfy one of the key decisive requirements of the (anti-) BRST symmetry 
transformations (connected with a gauge transformation) because 
the following is {\it not} true, namely;   
\begin{eqnarray}
(\tilde s_b \; \tilde s_{ab} + \tilde s_{ab} \; \tilde s_b ) \; \Psi = 0 
\hskip 1cm \Psi =  A_\mu, \; C, \; \bar C, \; B_{\mu \nu}, \; C_\mu, \; \bar C_\mu, \; \beta, 
\; \bar \beta, \; \phi,    
\end{eqnarray}
for the generic field $\Psi$ of the theory. For instance, it 
can be explicitly checked that we have the following relationships, namely;
\begin{eqnarray}
(\tilde s_b \; \tilde s_{ab} + \tilde s_{ab} \; \tilde s_b )\; C_\mu  \; = - \; \Box C_\mu \ne 0,\nonumber\\
(\tilde s_b \; \tilde s_{ab} + \tilde s_{ab} \; \tilde s_b )\; \bar C_\mu  \; = + 
\; \Box \bar C_\mu \ne 0.   
\end{eqnarray}
Thus, the nilpotent symmetry transformations $\tilde s_{(a)b}$ do not fulfill one 
of the central criteria of the BRST formalism. To obtain the off-shell nilpotent 
and absolutely anticommuting (anti-) BRST symmetry transformations, we shall take
recourse to the superfield formalism in the next section. \\

\noindent
{\bf 4. Off-shell nilpotent and absolutely anticommuting (anti-) BRST

symmetry transformations: superfield formalism}\\

\noindent
It is clear, from our earlier discussions, that the celebrated 4D topological term
(i.e. $(m/4) \varepsilon^{\mu\nu\eta\kappa} B_{\mu\nu} F_{\eta\kappa}$)
is a gauge (and, therefore, (anti-) BRST) invariant quantity. As a consequence,
for all practical purposes, the Lagrangian density ${\cal L}_{0}$ can be treated
as the sum of the {\it free} Abelian 1-form and 2-form gauge theories 
(whose nilpotent symmetries we are going to discuss below).

The off-shell nilpotent and absolutely anticommuting (anti-) BRST symmetry
transformations can be derived by exploiting the standard techniques of the
superfield formalism (see, e.g. [30,31] and [36-38] for details). For this paper
to be self-contained, we provide {\it firstly} a very concise description of
the superfield formalism, applied to the case of Abelian 1-form gauge theory [36-38]
(later on, we shall provide the superfield description of 2-form theory). 
In this context, it is worthwhile to 
point out that the curvature tensor $F_{\mu\nu}$, owing
its origin to the exterior derivative $d$ (i.e. $d A^{(1)} = (1/ 2!)
(dx^\mu \wedge dx^\nu) F_{\mu\nu}$), remains invariant under the (anti-) BRST
symmetry transformations. This observation remains intact as we 
proceed ahead from the ordinary 4D field theory to the superfield formalism
on the (4, 2)-dimensional supermanifold. Thus, first of all, we generalize 
(in our superfield formalism) the exterior derivative
$d$ to its counterpart on the (4, 2)-dimensional supermanifold as
\begin{eqnarray}
d \longrightarrow  \tilde d \; = \; d Z^M \partial_M \; \equiv \;  dx^\mu \; \partial_\mu  
+  d \theta \; \partial_\theta
+ d \bar \theta \; \partial_{\bar\theta},
\end{eqnarray}
where $Z^M = (x^\mu, \theta, \bar\theta),\; \partial_M = (\partial_\mu, \partial_\theta,
\partial_{\bar\theta}) $
are the superspace variables and corresponding partial derivatives on the
(4, 2)-dimensional supermanifold. Here the bosonic spacetime variables $x^\mu (\mu = 0, 1, 2, 3)$
and a pair of Grassmannian variables $\theta$ and $\bar\theta$
(with $\theta^2 = \bar\theta^2 = 0, \theta \bar\theta + \bar\theta \theta = 0$)
parametrize the above supermanifold. After this, we generalize the basic fields
$(A_\mu, C , \bar C)$, defined on the 4D ordinary spacetime Minkowski manifold, to the
corresponding superfields (defined on the (4, 2)-dimensional supermanifold) with the 
following expansions along the Grassmannian directions (see, e.g. [30-39])
\begin{eqnarray}
{\cal B}_\mu (x, \theta, \bar\theta) & = &
A_\mu (x) + \theta \; \bar R_\mu (x) + \bar\theta \; R_\mu (x) 
+ i \; \theta \; \bar \theta \; S_\mu (x),\nonumber\\
{\cal F} (x, \theta, \bar\theta) & = &
C (x) + i \; \theta \; \bar b_1 (x) 
+ i\;\bar\theta \; b_2 (x) + i \; \theta \; \bar \theta \; s (x),
\nonumber\\
\bar {\cal F} (x, \theta, \bar\theta) & = &
\bar C (x) + i \; \theta \; \bar b_2 (x) + i \; \bar\theta \; b_1 (x) 
+ i\;  \theta \; \bar\theta \; \bar s (x),
\end{eqnarray}
where, on the r.h.s., the fields $(R_\mu, \bar R_\mu, s , \bar s)$ 
and $(S_\mu, b_1, \bar b_1, b_2, \bar b_2)$
are the fermionic and bosonic secondary fields, respectively. These fields can be expressed
in terms of the basic and auxiliary fields of the theory when we exploit the potential
of the horizontality condition (HC).

The celebrated HC requires that the 2-form super-curvature should be equated with the 
ordinary 2-form curvature as follows
\begin{eqnarray}
\tilde d \;\tilde A^{(1)} = d\; A^{(1)} \; \Rightarrow 
\tilde F_{\mu\nu} (x,\theta,\bar\theta) = F_{\mu\nu} (x),
\end{eqnarray} 
where the super 1-form connection $\tilde A^{(1)}$ is defined, 
in terms of multiplet superfields
($ {\cal B}_\mu (x, \theta, \bar\theta), 
{\cal F} (x,\theta,\bar\theta), \bar {\cal F} (x, \theta, \bar\theta)$), as given below
\begin{eqnarray}
\tilde A^{(1)} = d Z^M A_M \equiv dx^\mu \; {\cal B}_\mu (x,\theta,\bar\theta) 
+ d \theta \; \bar {\cal F} (x,\theta,\bar\theta) + d \bar \theta \; 
{\cal F} (x,\theta,\bar\theta).
\end{eqnarray}
Furthermore, the HC (cf. (21)) also implies that the super-curvature tensor 
$\tilde F_{\mu\nu} (x,\theta,\bar\theta)$ is restricted to be equal to 
the ordinary curvature tensor $F_{\mu\nu} (x)$.
The above restriction (i.e. HC) yields the following relationships [39]
\begin{eqnarray}
&& b_2 = \bar b_2 = 0, \qquad s  = \bar s = 0, \qquad b_1 + \bar b_1 = 0, \nonumber\\
&& R_\mu = \partial_\mu C, \qquad \bar R_\mu = \partial_\mu \bar C, 
\qquad S_\mu = \partial_\mu B,
\end{eqnarray}
where we have chosen the secondary fields  $b_1$ and $\bar b_1$ in terms
of the Nakanishi-Lautrup auxiliary field $B$ (i.e.  $b_1 = B = - \; \bar b_1$). The latter
is required to linearize the gauge-fixing term (i.e. $B (\partial \cdot A) + (1/2) B^2
= - (1/2) (\partial \cdot A)^2$) in the ordinary
(anti-) BRST invariant Lagrangian density (see, Sec. 5 below). 
Substitution of these fields in the superfield expansions yields the following
\begin{eqnarray}
{\cal B}^{(h)}_\mu (x, \theta, \bar\theta) & = &
A_\mu (x) + \theta\; (\partial_\mu \bar C (x)) + \bar \theta\; (\partial_\mu C (x)) 
+ i\;\theta\;\bar\theta\; (\partial_\mu B (x)) \nonumber\\ 
&\equiv& A_\mu (x) + \theta \; (s_{ab} A_\mu (x)) + \bar\theta \;  
(s_b A_\mu (x)) + \theta \; \bar \theta \; (s_b s_{ab} A_\mu (x)),\nonumber\\
{\cal F}^{(h)} (x, \theta, \bar\theta) & = &
C (x) - i \; \theta \; B (x) \; \equiv \;  C (x) + \theta \; (s_{ab} C (x)), \nonumber\\
\bar {\cal F}^{(h)} (x, \theta, \bar\theta) & = &
\bar C (x) + i  \; \bar\theta \; B (x) \; \equiv \; \bar C (x) + \bar \theta \; (s_b \bar C (x)),
\end{eqnarray}
where the superscript $(h)$ stands for the superfield expansion
after the application of the HC and we have denoted the off-shell nilpotent and absolutely 
anticommuting (anti-) BRST symmetry transformations as $s_{(a)b}$\footnote {In explicit 
terms, it can be seen that we have derived: $s_b A_\mu = \partial_\mu C,\; s_b C = 0,
\; s_b \bar C = i B, \; s_b B = 0$ and $s_{ab} A_\mu = \partial_\mu \bar C, \;
s_{ab} \bar C = 0, \;   s_{ab} C = - i B, \; s_{ab} B = 0.$}.

In exactly above fashion, we can {\it now} generalize the basic fields $B_{\mu\nu}, C_\mu, 
\bar C_\mu,$ $\phi, \beta, \bar\beta$ of the ordinary 4D Abelian 2-form gauge theory
onto the (4, 2)-dimensional supermanifold and these superfields would
have the expansions along the Grassmannian directions as (see, [27] for details)
\begin{eqnarray}
{\cal B}_{\mu\nu} (x, \theta, \bar\theta) &=& B_{\mu\nu}
(x) + \theta\; \bar R_{\mu\nu} (x) + \bar\theta\; R_{\mu\nu} (x) +
i \;\theta \; \bar\theta\; S_{\mu\nu} (x), \nonumber\\ 
{\cal F}_\mu (x, \theta, \bar\theta) &=& C_\mu
(x) + \theta \;\bar B^{(1)}_\mu (x) + \bar\theta\; B^{(1)}_\mu (x)
+ i \;\theta \; \bar\theta\;f^{(1)}_\mu (x), \nonumber\\ 
{\bar {\cal F}}_\mu (x, \theta, \bar\theta) &=& \bar C_\mu (x) +
\theta\; \bar B^{(2)}_\mu (x) + \bar\theta\; B^{(2)}_\mu (x) + i
\; \theta\; \bar\theta \bar f^{(2)}_\mu (x), \nonumber\\
\beta (x, \theta, \bar\theta ) &=& \beta (x) + \theta \;\bar f_1 (x) +
\bar\theta\; f_1 (x) + i\; \theta\; \bar\theta\; b_1 (x),\nonumber\\ 
{\bar \beta} (x, \theta, \bar\theta) &=& 
\bar\beta (x) + \theta \;\bar f_2 (x) + \bar \theta\; f_2 (x) + i
\;\theta\;\bar\theta\; b_2 (x), \nonumber\\
\Phi (x,\theta, \bar\theta) &=& \phi (x) + \theta \;\bar f_3 (x) +
\bar\theta\; f_3 (x) + i \;\theta \;\bar\theta\; b_3 (x),
\end{eqnarray} 
where ($R_{\mu\nu}, \bar R_{\mu\nu}, f_1, \bar f_1, f_2, \bar f_2,
f_3, \bar f_3, f_\mu^{(1)}, \bar f_\mu^{(2)}$) and 
($ S_{\mu\nu}, B^{(1)}_\mu, \bar B^{(1)}_\mu, B_\mu^{(2)}, \bar B_\mu^{(2)},$
$ b_1, b_2, b_3$) are the fermionic and bosonic set of secondary fields, respectively.
In terms of the above superfields, the super 2-form connection 
on the (4, 2)-dimensional supermanifold can be written as (see, e.g. [27])
\begin{eqnarray}
\tilde B^{(2)} &=& {\displaystyle \frac{1}{2!}} (dx^\mu \wedge
dx^\nu) \; {\cal B}_{\mu\nu} (x, \theta, \bar\theta)\;
+ \; (dx^\mu \wedge d \theta)\; \bar {\cal F}_\mu (x, \theta, \bar\theta) \nonumber\\
&+& (dx^\mu \wedge d \bar\theta)\; {\cal F}_\mu (x, \theta, \bar\theta)\; 
+ \; (d\theta \wedge d \theta)\; {\bar \beta} (x, \theta, \bar\theta) \nonumber\\ 
&+& (d \bar\theta \wedge d \bar\theta) \; \beta (x, \theta, \bar\theta)\; 
+ \; (d \theta \wedge d\bar\theta)\; \Phi  (x, \theta, \bar\theta).
\end{eqnarray}
The celebrated HC for this system can be expressed as 
\begin{equation}
\tilde d \; \tilde B^{(2)} \; = \; d \; B^{(2)} \; \Longrightarrow \;
\tilde H_{\mu\nu\eta} (x, \theta, \bar\theta) = H_{\mu\nu\eta} (x).
\end{equation} 
In other words, the HC is a restriction such that the super-curvature
tensor $\tilde H_{\mu\nu\eta} (x,\theta,\bar\theta)$ is, ultimately,
independent of the Grassmannian variables so that 
$\tilde H_{\mu\nu\eta} (x, \theta, \bar\theta) = H_{\mu\nu\eta} (x)$.
The above condition leads to the following relationships amongst the
basic, secondary and auxiliary fields [27]
\begin{eqnarray}
&b_1 =  b_2 = b_3 = 0, \quad f_1 = 0, \quad \bar f_2 = 0, 
\quad \bar f_1 + f_3 = 0, \quad f_2 + \bar f_3 = 0, &\nonumber\\
& \bar B_\mu^{(1)} + B_\mu^{(2)} + \partial_\mu \phi \; = \; 0, 
\quad B_\mu^{(1)} \; = \; - \; \partial_\mu \beta, 
\quad \bar B_\mu^{(2)} \; = \;  - \; \partial_\mu \bar \beta , &\nonumber\\
& f_\mu^{(1)} \; = \;  i \; \partial_\mu f_3 \; \equiv \; 
- \; i \; \partial_\mu \bar f_1, \quad
\bar f_\mu^{(2)} \; = \;  - \; i \; \partial_\mu \bar f_3 \; \equiv \; 
+ \; i \; \partial_\mu f_2, &\nonumber\\
& R_{\mu\nu} \; = \;  - \; (\partial_\mu C_\nu - \partial_\nu C_\mu), \quad \bar R_{\mu\nu} 
\; = \; - \;(\partial_\mu \bar C_\nu - \partial_\nu \bar C_\mu), &\nonumber\\
& S_{\mu\nu} \; = \; - \; i \; (\partial_\mu \bar B_\nu - \partial_\nu \bar B_\mu) 
\; \equiv \; - \; i \; (\partial_\mu B_\nu - \partial_\nu B_\mu).& 
\end{eqnarray}
We can make the following choices for the algebraic convenience:
\begin{eqnarray}
&& \bar f_3 \; = \; \rho (x) \; = \; - f_2 (x), \qquad \bar B_\mu^{(1)} \; = \; \bar B_\mu, \nonumber\\
&& f_3 \; = \; \lambda (x) \; = \; - \bar f_1 (x), \qquad B_\mu^{(2)} \; = \; - B_\mu,
\end{eqnarray}
which lead to the derivation of a Curci-Ferrari (CF)-type restriction, in  the realm
of the Abelian 2-form gauge theory, as:
\begin{eqnarray}
\bar B_\mu^{(1)} + B_\mu^{(2)} + \partial_\mu \phi \; = \; 0 \; \Longrightarrow \;
B_\mu - \bar B_\mu - \partial_\mu \phi = 0.
\end{eqnarray}
The above condition is responsible for the absolute anticommutativity of the (anti-)
BRST symmetry transformations as we elaborate below.

After the substitution of the expressions for the secondary fields, the explicit expansions
for the superfields are 
\begin{eqnarray}
{\cal B}^{(h)}_{\mu\nu} (x, \theta, \bar\theta) &=& B_{\mu \nu} (x) 
- \theta \; (\partial_\mu \bar C_\nu - \partial_\nu \bar C_\mu) 
- \bar \theta \; (\partial_\mu C_\nu - \partial_\nu C_\mu)\nonumber\\
&+& \theta \; \bar \theta \; (\partial_\mu B_\nu - \partial_\nu B_\mu)\nonumber\\
&\equiv& B_{\mu\nu} (x) + \theta\; (s_{ab} \; B_{\mu\nu} (x)) + \bar\theta\;
(s_b \; B_{\mu\nu} (x)) \nonumber\\
&+& \theta\; \bar\theta\; (s_b \; s_{ab} \; B_{\mu\nu} (x)), \nonumber\\
{\cal F}^{(h)}_\mu (x, \theta, \bar\theta) &=& C_\mu (x) + \theta \; \bar B_{\mu} (x) 
- \bar \theta \; \partial_\mu \beta - \theta \; \bar\theta \; \partial_\mu \lambda \nonumber\\
&\equiv& C_\mu (x) + \theta \;(s_{ab} \; C_\mu (x)) +
\bar\theta\; (s_b \; C_\mu (x)) \nonumber\\
&+& \theta\; \bar\theta\; (s_b \; s_{ab} \; C_\mu (x)), \nonumber\\ 
{\bar {\cal F}}^{(h)}_\mu (x, \theta, \bar\theta) &=& \bar C_\mu (x) 
- \theta \; \partial_{\mu} \bar \beta  - \bar \theta \; B_\mu (x)  
+ \theta \; \bar\theta \; \partial_\mu \rho \nonumber\\
&\equiv & \bar C_\mu (x) + \theta\; (s_{ab} \; \bar
C_\mu (x)) + \bar\theta\; (s_b \; \bar C_\mu (x)) \nonumber\\
&+& \theta\; \bar\theta\;  (s_b \; s_{ab} \; \bar C_\mu(x)),\nonumber\\
\Phi^{(h)} (x, \theta, \bar\theta) &=& \phi (x) + \theta \; \rho (x)
+ \bar \theta \; \lambda (x) \nonumber\\
&\equiv& \phi (x) + \theta \;(s_{ab} \; \phi (x)) + \bar\theta\; (s_b \; \phi (x)), \nonumber\\
\beta^{(h)} (x, \theta, \bar\theta ) &=& \beta (x) - \theta \; \lambda (x)
\equiv \beta (x) + \theta \;(s_{ab}\; \beta (x)), \nonumber\\ 
{\bar \beta}^{(h)} (x, \theta, \bar\theta) &=&  \bar \beta (x) - \bar \theta \; \rho(x)
\equiv  \bar\beta (x) + \bar \theta\; (s_b \;\bar \beta (x)), 
\end{eqnarray}
where the superscript $(h)$ denotes the superfield expansion after the application of HC.
The above expansions yield the following off-shell nilpotent (anti-) BRST symmetry
transformations for the relevant fields of the theory
\begin{eqnarray}
&& s_b B_{\mu \nu} = - \; (\partial_\mu C_\nu - \partial_\nu C_\mu), \quad s_b C_\mu 
= - \; \partial_\mu \beta, \quad s_b \bar C_\mu = - \; B_\mu, \nonumber\\
&& s_b \bar \beta = - \; \rho, \quad s_b \phi = \lambda, \quad s_b \bigl[\rho,\; \lambda,\;
\beta, \; B_\mu, \; H_{\mu \nu \eta} \bigr] = 0,
\end{eqnarray} 
\begin{eqnarray}
&& s_{ab} B_{\mu \nu} = - \; (\partial_\mu \bar C_\nu - \partial_\nu \bar C_\mu), 
\quad s_{ab} \bar C_\mu 
= - \; \partial_\mu \bar \beta, \quad s_{ab} C_\mu = + \; \bar B_\mu, \nonumber\\
&& s_{ab} \beta = - \; \lambda, \quad s_{ab} \phi = \rho, \quad s_b \bigl[\rho,\; \lambda,\;
\bar \beta, \; \bar B_\mu, \; H_{\mu \nu \eta} \bigr] = 0.
\end{eqnarray}
The absolute anticommutativity requirement imposes the (anti-) BRST symmetry transformations
on the Nakanishi-Lautrup type auxiliary fields as:
\begin{eqnarray} 
s_b \bar B_\mu = -\partial_\mu \lambda, \quad s_{ab} B_\mu = \partial_\mu \rho.    
\end{eqnarray}
Thus, we have obtained the complete set of (anti-) BRST symmetry transformations in the
equations (32), (33) and (34) which are off-shell nilpotent of order two and they are
absolutely anticommuting in nature as can be checked from the following explicit example:
\begin{eqnarray} 
\{ s_b, \; s_{ab} \} \; B_{\mu \nu} = \partial_\mu \bigl(B_\nu - \bar B_\nu \bigr) 
- \partial_\nu \bigl(B_\mu - \bar B_\mu \bigr) = 0.
\end{eqnarray}
The r.h.s. of the above equation is zero on the constrained surface defined by 
the equation (30) (which is nothing but the CF-type restriction). For the rest of the 
fields of the theory, it can be checked that $\{ s_b, \; s_{ab} \} \Psi = 0 $ 
for $\Psi$ being the generic field (except $B_{\mu\nu}$ that has been considered in (35)).\\

\noindent
{\bf 5. Nilpotent symmetry invariance: Lagrangian formalism}\\

\noindent
We begin with the BRST and anti-BRST invariant coupled Lagrangian densities,
corresponding to the starting Lagrangian density (1), as  
\begin{eqnarray}
{\cal L}_B &=& - \; \frac {1}{4}\; F^{\mu \nu} F_{\mu \nu} + \frac {1}{12} \; H^{\mu \nu \eta}
H_{\mu \nu \eta} + \frac {1}{4} \; m \; \varepsilon^{\mu \nu \eta \kappa}
B_{\mu \nu} F_{\eta \kappa} + B (\partial \cdot A) \nonumber\\
&+& \frac {1}{2} \; B^2 + B^\mu (\partial^\nu B_{\nu \mu} - \partial_\mu \phi) + B\cdot B
- i \; \partial_\mu \bar C \; \partial^\mu C 
+  \partial_\mu \bar \beta \; \partial^\mu \beta \nonumber\\
&+& (\partial_\mu \bar {C_\nu} - \partial_\nu \bar {C_\mu}) \; (\partial^\mu C^\nu)
+ (\partial \cdot C - \lambda )\rho + (\partial \cdot \bar C + \rho ) \lambda,  
\end{eqnarray}
\begin{eqnarray}
{\cal L}_{\bar B} 
&=& - \; \frac {1}{4} \; F^{\mu \nu} F_{\mu \nu} + \frac {1}{12}\; H^{\mu \nu \eta}
H_{\mu \nu \eta} + \frac {1}{4} \; m \; \varepsilon^{\mu \nu \eta \kappa}
B_{\mu \nu} F_{\eta \kappa} + B (\partial \cdot A) \nonumber\\
&+& \frac {1}{2} \; B^2 + \bar B^\mu (\partial^\nu B_{\nu \mu} + \partial_\mu \phi) 
+ \bar B\cdot \bar B
- i \; \partial_\mu \bar C \; \partial^\mu C 
+  \partial_\mu \bar \beta \; \partial^\mu \beta \nonumber\\
&+& (\partial_\mu \bar {C_\nu} - \partial_\nu \bar {C_\mu}) \; (\partial^\mu C^\nu)
+ (\partial \cdot C - \lambda )\rho + (\partial \cdot \bar C + \rho ) \lambda,
\end{eqnarray}
where the scalar field $B$ and vector fields $(B_\mu, \bar B_\mu)$ are the Nakanishi-Lautrup
type auxiliary fields, the scalar $(\bar C, C)$ and vector $(\bar C_\mu, C_\mu)$ 
fields are the fermionic (anti-) ghost fields, $(\bar \beta, \beta)$ are the 
bosonic ghost for ghost fields, $(\rho) \lambda$ are the fermionic auxiliary 
(anti-) ghost fields and the massless $(\Box \phi = 0)$ scalar field $\phi$ 
is required in the gauge-fixing term for the stage-one reducibility (that is 
present in the second-rank antisymmetric tensor gauge theory).

The Lagrangian density $({\cal L}_B)$ respects the following off-shell nilpotent 
$(s^2_b = 0)$ BRST symmetry\footnote{These transformations and (43)
(see below) have been obtained (cf. (32), (33))
by exploiting the superfield approach to BRST formalism in the context of Abelian 2-form
gauge theory in our previous section (see, [27] for details). We take here  an overall 
minus sign so that we could be consistent with the transformations in Secs. 2 and 3
for the sake of precise comparison (at least, for the gauge and (anti-) BRST 
transformations on $B_{\mu\nu}$).} transformations $(s_b)$ [27]
\begin{eqnarray}
&&s_b A_\mu = \partial_\mu C, \quad s_b \bar C = i B, \quad 
s_b B_{\mu \nu} = (\partial_\mu C_\nu - \partial_\nu C_\mu), \nonumber\\
&& s_b C_\mu = \partial_\mu \beta, \quad s_b \bar C_\mu = B_\mu, 
\quad s_b \phi =  - \lambda, \quad s_b \bar \beta = \rho, \nonumber\\
&& s_b [C, \; B, \; \rho, \; \lambda, \; \beta, \;  B_\mu, \; H_{\mu \nu \kappa}] = 0,   
\end{eqnarray}
because the above Lagrangian density transforms to a total spacetime derivative as
given below:
\begin{eqnarray}
s_b {\cal L}_B &=& \partial_\mu \Bigl[B \; \partial^\mu C + \rho \; \partial^\mu \beta
+ \lambda \; B^\mu + (\partial^\mu C^\nu - \partial^\nu C^\mu )\; B_\nu \nonumber\\
&+& m \; \varepsilon ^{\mu \nu \eta \kappa} C_\nu \; (\partial_\eta A_\kappa) \Bigr]. 
\end{eqnarray}
As a consequence, the action $S_{(B)} = \int d^4 x \; {\cal L}_B$ remains invariant 
under the off-shell nilpotent BRST symmetry transformations $(s_b)$.

The Noether conserved current $(J^\mu_{(B)})$, that emerges due to 
the continuous BRST symmetry transformations $(s_b)$, is   
\begin{eqnarray}
J^\mu_{(B)} &=& (\partial^\mu C^\nu - \partial^\nu C^\mu ) \; B_\nu
+ \frac {1}{2} \; m \; \varepsilon ^{\mu \nu \eta \kappa} (\partial_\nu C) \;  B_{\eta \kappa}
+ H^{\mu \nu \eta} (\partial_\nu C_\eta )\nonumber\\
&+& B \; \partial^\mu C - F^{\mu \nu} (\partial_\nu C) - (\partial^\mu \bar C^\nu
- \partial^\nu \bar C^\mu ) \; (\partial_\nu \beta ) + \lambda \; B^\mu  \nonumber\\
&+& \rho \; \partial^\mu \beta - m \; \varepsilon ^{\mu \nu \eta \kappa} 
C_\nu (\partial_\eta A_\kappa).
\end{eqnarray}
The conservation law $(\partial_\mu J^\mu_{(B)} = 0)$ can be proven by exploiting 
the following equations of motion that emerge from ${\cal L}_B$:
\begin{eqnarray}
&& \partial_\mu H^{\mu \nu \eta} + (\partial^\nu B^\eta - \partial^\eta B^\nu )
- \frac {1}{2} \; m \; \varepsilon^{\nu \eta \kappa \zeta} F_{\kappa \zeta} = 0, \qquad
(\partial \cdot B) = 0, \nonumber\\
&& B_\mu = - \frac {1}{2} \; (\partial^\nu B_{\nu \mu} - \partial_\mu \phi ), \quad
\partial_\mu F^{\mu \nu} = \partial^\nu B - \frac {1}{2} \; m \; 
\varepsilon^{\nu \mu \kappa \eta} \; (\partial_\mu B_{\kappa \eta}), \nonumber\\
&& \Box \beta = 0, \quad \Box \bar \beta = 0, \quad  
\Box C = 0, \quad  \Box \bar C = 0, \quad B = - \;(\partial \cdot A),\nonumber\\
&&\Box \phi = 0, \qquad \lambda = + \;\frac{1}{2} \;(\partial \cdot C), \qquad
\rho = -\; \frac{1}{2}\; (\partial \cdot \bar C), \nonumber\\
&&\Box C_\mu = \partial_\mu \lambda 
\; \equiv \; \frac {1}{2} \; \partial_\mu (\partial \cdot C), \quad \Box \bar C_\mu 
= - \partial_\mu \rho \; \equiv \; \frac {1}{2} \; \partial_\mu (\partial \cdot \bar C).
\end{eqnarray}
The conserved BRST charge $Q_B$, corresponding to $J^\mu_{(B)}$ would be given by 
$Q_B = \int d^3 x \;J^0_{(B)}$, whose explicit form is:
\begin{eqnarray}
Q_{(B)} &=& \int d^3 x \Bigl[B \dot C - \dot B C + \Pi^{i j} \; (\partial_i C_j 
- \partial_j C_i) + (\partial^0 C^i - \partial^i C^0) B_i \nonumber\\
&-& (\partial^0 \bar C^i - \partial^i \bar C^0) \; (\partial_i \beta ) + \rho \; \dot \beta
+ \lambda \; B_0 + m \; \epsilon_{i j k} \; C_i \; (\partial_j A_k)  \Bigr ].
\end{eqnarray}
It can be checked that it is a conserved ($\dot Q_{(B)} = 0$) and nilpotent 
($Q_{(B)}^2 = 0$). A close look at $Q_{(B)}$ ensures that it is a generalization
of the expressions in (7) and (9) (cf. Sec. 2) for $Q_{(gt)}$ and $G$, respectively.

The Lagrangian density ${\cal L}_{\bar B}$ respects the following 
off-shell nilpotent $(s^2_{ab} = 0)$ anti-BRST symmetry transformations $(s_{ab})$
\begin{eqnarray}
&&s_{ab} A_\mu = \partial_\mu \bar C, \quad s_{ab} C = - i B, \quad 
s_{ab} B_{\mu \nu} = (\partial_\mu \bar C_\nu - \partial_\nu \bar C_\mu), \nonumber\\
&& s_{ab} \bar C_\mu = \partial_\mu \bar \beta, \quad s_{ab} C_\mu = - \bar B_\mu, \quad 
s_{ab} \phi =  - \rho, \quad s_{ab} \beta = \lambda, \nonumber\\
&&s_{ab} [\bar C, \; B, \;  \rho, \; \lambda, \; \bar \beta, \; \bar B_\mu, \; 
H_{\mu \nu \kappa}] = 0, 
\end{eqnarray}
because the above ${\cal L}_{\bar B}$ transforms to a total spacetime 
derivative\footnote {Under the BRST symmetry transformations
$s_b$ (with $s_b \bar B_\mu = \partial_\mu \lambda $ ), the Lagrangian density 
${\cal L}_{\bar B}$ transforms to a total spacetime derivative plus a term that is zero on 
the constrained surface defined by $B_\mu - \bar B_\mu = \partial_\mu \phi$. Exactly, in
a similar fashion, ${\cal L}_B$ transforms under $s_{ab}$ (with 
$s_{ab} B_\mu = - \partial_\mu \rho$) to a total spacetime 
derivative plus a term that is zero on the constrained surface defined by 
field equation $B_\mu - \bar B_\mu = \partial_\mu \phi$.}
\begin{eqnarray}
s_{ab} {\cal L}_{\bar B} &=& \partial_\mu \Bigl[B \; \partial^\mu \bar C - \rho \; \bar B^\mu 
+ \lambda \; \partial^\mu \bar \beta + (\partial^\mu \bar C^\nu 
- \partial^\nu \bar C^\mu ) \; \bar B_\nu \nonumber\\
&+& m \; \varepsilon ^{\mu \nu \eta \kappa} \bar C_\nu \; (\partial_\eta A_\kappa) \Bigr]. 
\end{eqnarray} 
As a result, the action $(S_{(\bar B)} = \int d^4 x \; {\cal L}_{\bar B})$ remains invariant.
It should be noted that ${\cal L}_B$ and ${\cal L}_{\bar B}$ are equivalent due to 
$B_\mu - \bar B_\mu - \partial_\mu \phi = 0$.

The symmetry invariance under the continuous nilpotent transformations $s_{ab}$ 
implies a Noether's conserved current as given by
\begin{eqnarray}
J^\mu_{(\bar B)} &=& (\partial^\mu \bar C^\nu - \partial^\nu \bar C^\mu ) \;\bar B_\nu
+ \frac {1}{2} \; m \; \varepsilon^{\mu \nu \eta \kappa} B_{\nu \eta} \; 
(\partial_\kappa \bar C) + H^{\mu \nu \eta} (\partial_\nu \bar C_\eta ) \nonumber\\
&+& B \; \partial^\mu \bar C - F^{\mu \nu} (\partial_\nu \bar C) + (\partial^\mu C^\nu 
- \partial^\nu C^\mu ) \; (\partial_\nu \bar \beta ) - \rho \; \bar B^\mu \nonumber\\
&+& \lambda \; \partial^\mu \bar \beta - m \; \varepsilon ^{\mu \nu \eta \kappa} \bar C_\nu 
(\partial_\eta A_\kappa).
\end{eqnarray}  
The conservation law $(\partial_\mu J^\mu_{(\bar B)} = 0)$ can be proven by taking into 
account the equations of motion from ${\cal L}_{(\bar B)}$ that are same as (41) except 
the following:
\begin{eqnarray}
&& \bar B_\mu = - \; \frac{1}{2} \;(\partial^\nu B_{\nu\mu} + \partial_\mu \phi), \qquad 
(\partial \cdot \bar B) = 0 \; \Longrightarrow  \; \Box \phi = 0, \nonumber\\
&& \partial_\mu H^{\mu \nu \eta} \;+\; (\partial^\nu \bar B^\eta \;- \;\partial^\eta \bar B^\nu )
- \frac {1}{2} \; m \; \varepsilon^{\nu \eta \kappa \zeta} F_{\kappa \zeta} = 0.
\end{eqnarray}
The corresponding anti-BRST charge ($Q_{(\bar B)} = \int d^3 x \; J^0_{(\bar B)}$) is 
\begin{eqnarray}
Q_{(\bar B)} 
&=& \int d^3 x \Bigl[ B \dot {\bar C} - \dot B \bar C + \Pi^{i j} \;(\partial_i \bar C_j 
- \partial_j \bar C_i) + (\partial^0 C^i - \partial^i C^0) \bar B_i \nonumber\\
&+& (\partial^0 C^i - \partial^i C^0) \; (\partial_i \bar \beta ) - \rho \; \bar B_0
+ \lambda \; \dot { \bar \beta} 
+ m \; \epsilon_{i j k} \; \bar C_i \; (\partial_j A_k)\Bigr ].
\end{eqnarray}
The above charge is also a conserved ($\dot Q_{(\bar B)} =  0$) and nilpotent
($Q_{(\bar B)}^2 = 0$). Like $Q_{(B)}$, the anti-BRST charge $Q_{(\bar B)}$ is
also generalization of (7) and (9).

It is worth noting that the equations of motion (41) and (46) imply that the
relationship in equation (30) (that corresponds to CF condition) is true. In fact, 
this constrained field equation defines a surface on the 4D-spacetime 
manifold where the absolute anticommutativity of the (anti-) BRST symmetries is
satisfied. This relationship has also been shown to be connected with the geometrical 
object called gerbs which are one of the very active areas of research in theoretical 
high energy physics [40,41].
Now we dwell a bit on the conditions that emerge from the physicality criteria 
$Q_{(\bar B) B} |\; phys > \; = 0$. It can be seen that the conserved and nilpotent 
BRST charge $Q_{(B)}$ produces   
\begin{eqnarray}
Q_{(B)} |\; phys > \; = \; 0\Longrightarrow \Pi^0_{(A)} |\; phys > \; = \; 0 
\quad \Longrightarrow B |\; phys > \; = \; 0, \nonumber\\ 
 \partial_i \Pi_{(A)}^i |\; phys > \; = \; 0 
\quad \Longrightarrow \dot B |\; phys > \; = \; 0, \nonumber\\
\Pi^{0 i}_{(B)}|\; phys > \; = \; 0 \quad \Longrightarrow B^i |\; phys > \; = \; 0, \nonumber\\
\partial_i \Pi^{i j}_{(B)} |\; phys > \; = \; 0 
\quad \Longrightarrow \partial_i H^{0 i j} |\; phys > \; = \; 0.   
\end{eqnarray}
The same conditions also emerge from the anti-BRST charge 
$Q_{(\bar B)} |\; phys > \; = 0$. The above condition (48) ensure that the BRST 
quantization method is consistent with the requirements of the Dirac's method of 
quantization of systems with constraints. Thus, the BRST quantization scheme 
resolves all the unanswered issues that were raised at the fag end of Sec. 2.\\

\noindent
{\bf 6. (Anti-) BRST invariance: superfield formalism}\\

\noindent
It is interesting to point out that the coupled Lagrangian densities (36) and (37) can be 
expressed (modulo some total spacetime derivatives) as 
\begin{eqnarray}
{\cal L}_B &=& - \; \frac {1}{4}\; F^{\mu \nu} F_{\mu \nu} + \frac {1}{12} \; H^{\mu \nu \eta}
H_{\mu \nu \eta} + \frac {1}{4} \; m \; \varepsilon^{\mu \nu \eta \kappa} 
B_{\mu \nu} F_{\eta \kappa}\nonumber\\
&+& s_b \; s_{ab} \Bigl[\frac {i}{2}\; A_\mu A^\mu + \; \frac {1}{2}\; C \bar C
- \; \frac {1}{4}\; B_{\mu \nu} B^{\mu \nu} + \bar C_\mu C^\mu + 2 \beta \bar \beta \Bigr ],  
\end{eqnarray}      
\begin{eqnarray}
{\cal L}_{\bar B} &=& - \; \frac {1}{4}\; F^{\mu \nu} F_{\mu \nu} 
+ \frac {1}{12} \; H^{\mu \nu \eta}
H_{\mu \nu \eta} + \frac {1}{4} \; m \; \varepsilon^{\mu \nu \eta \kappa} 
B_{\mu \nu} F_{\eta \kappa}\nonumber\\
&-& s_{ab} \; s_b \Bigl[\frac {i}{2}\; A_\mu A^\mu + \; \frac {1}{2}\; C \bar C
- \; \frac {1}{4}\; B_{\mu \nu} B^{\mu \nu} + \bar C_\mu C^\mu + 2 \beta \bar \beta \Bigr ],
\end{eqnarray}      
where we have to exploit the (anti-) BRST transformations quoted in (43) and (38).
Furthermore, we have to tap the usefulness of the CF-type restriction 
(that is written in (30)) so that ${\cal L}_B$ and ${\cal L}_{\bar B}$ can be 
expressed in the particular forms that are quoted in equations (36) and (37).

It is evident from the super expansion (24) and (31) that the (anti-) BRST 
symmetry transformations for a 4D ordinary field can be expressed in terms of the translations 
of the corresponding superfield along the Grassmannian directions of the 
(4, 2)-dimensional supermanifold, as\footnote{It should be noted that there is an overall
sign difference between the transformations ((32), (33)) and ((38), (43)). Thus,
the mapping quoted below (cf. (51)) is correct modulo the sign factor in the context of 
(anti-) BRST symmetries for the Abelian 2-form theory.}
\begin{eqnarray} 
s_b \Omega (x) = \lim_{\theta \to 0} \; \frac {\partial }{\partial \bar \theta}\; 
\Omega^{(h)} (x, \theta, \bar \theta), 
\quad s_{ab} \Omega (x) = \lim_{\bar \theta \to 0} \; \frac {\partial }{\partial \theta}
\; \Omega^{(h)} (x, \theta, \bar \theta),    
\end{eqnarray}
where $\Omega (x)$ is the generic 4D field and $\Omega^{(h)} (x, \theta, \bar \theta)$
is  the corresponding superfield. Furthermore, it is also clear from expansions (24)
and (31) that 
\begin{eqnarray}
\frac {\partial}{\partial \theta} \; \frac {\partial}{\partial \bar \theta} 
\; \Omega^{(h)} (x,\theta,\bar\theta) \;+ \;\frac {\partial}{\partial \bar \theta} 
\; \frac {\partial}{\partial \theta} \; \Omega^{(h)} (x,\theta,\bar\theta)\; = \;  0,
\end{eqnarray}
which corresponds to the anticommutativity of the (anti-) BRST symmetry transformations
in the operator form (cf. (17)). The above expressions provide the geometrical interpretations
for the (anti-) BRST symmetry transformations (and their corresponding generators) in
the language of the translational generators $(\partial_\theta, \partial_{\bar\theta})$
(with $\partial_\theta^2 = \partial_{\bar\theta}^2 = 0, 
\partial_\theta \partial_{\bar\theta} + \partial_{\bar \theta} \partial_{\theta} = 0$)
along the Grassmannian directions of the (4, 2)-dimensional supermanifold.

We have to recall that the HCs of (21) and (27) imply that 
\begin{eqnarray}
\tilde F_{\mu\nu}^{(h)} (x, \theta, \bar\theta) \; = \;  \partial_\mu \; {\cal B}_\nu^{(h)} 
(x, \theta, \bar \theta) - \partial_\nu \; {\cal B}_\mu^{(h)} 
(x, \theta, \bar \theta) \; = \; F_{\mu\nu} (x),
\end{eqnarray}
\begin{eqnarray}
\tilde H_{\mu\nu\eta}^{(h)} (x, \theta, \bar\theta) \; &=& \;  \partial_\mu \; 
{\cal B}_{\nu\eta}^{(h)} (x, \theta, \bar \theta) + \partial_\nu \; 
{\cal B}_{\eta\mu}^{(h)} (x, \theta, \bar \theta) + \partial_\eta \; 
{\cal B}_{\mu\nu}^{(h)} (x, \theta, \bar \theta) \nonumber\\
&=& H_{\mu\nu\eta} (x),
\end{eqnarray} 
where $\tilde F_{\mu\nu}^{(h)} (x, \theta, \bar\theta)$ and 
$\tilde H_{\mu\nu\eta}^{(h)} (x, \theta, \bar\theta)$ are the super-curvature
tensors after the application of the HC. The above conditions (53) and (54) 
show that $\tilde F_{\mu\nu}^{(h)} (x, \theta, \bar\theta)$ and 
$\tilde H_{\mu\nu\eta}^{(h)} (x, \theta, \bar\theta)$ are basically independent 
of the Grassmannian variables which can be proven by taking into account (24) and
(31). Thus, the kinetic term of the Lagrangian densities 
(49) and (50) can be written in terms of the superfields (after the application of HC) as:
\begin{eqnarray}
- \; \frac {1} {4} \; \tilde F^{\mu\nu (h)} (x, \theta, \bar\theta) \; 
\tilde F_{\mu\nu}^{(h)} (x, \theta, \bar\theta) + \frac {1} {12} \;
\tilde H^{\mu\nu\eta (h)} (x, \theta, \bar\theta) \; 
\tilde H_{\mu\nu\eta}^{(h)} (x, \theta, \bar\theta),
\end{eqnarray}
which are actually independent of Grassmannian variables. 
Thus, without the inclusion of the topological $ \bigl[T (x) = (B \wedge F) (x)\bigr]$ 
term, we can express the rest part of the Lagrangian densities (49) and (50), 
in the language of superfields (obtained after the application of HC), as
\begin{eqnarray}
\tilde {\cal L}_B &=& - \; \frac {1} {4} \; \tilde F_{\mu\nu}^{(h)} \; 
\tilde F^{\mu\nu (h)} + \frac {1} {12} \; \tilde H_{\mu\nu\eta}^ {(h)} \; 
\tilde H^{\mu\nu\eta (h)} + \frac {\partial}{\partial \bar \theta} \; 
\frac {\partial}{\partial \theta} \;\Bigl [ \frac {i}{2} \; {\cal B}_\mu^{(h)}
\; {\cal B}^{\mu{(h)}} \nonumber\\
&+& \; \frac {1}{2}\;  {\cal F}^{(h)} \;\bar {\cal F}^{(h)} 
- \; \frac {1}{4} \; {\cal B}_{\mu\nu} ^{(h)} \; {\cal B}^{\mu\nu (h)}
+ \bar {\cal F}_\mu^{(h)} {\cal F}^{\mu (h)} + 2\; \beta^{(h)} \; \bar \beta^{(h)} \Bigr ], 
\end{eqnarray}
\begin{eqnarray}
\tilde {\cal L}_{\bar B} &=& - \; \frac {1} {4} \; \tilde F_{\mu\nu}^{(h)} \; 
\tilde F^{\mu\nu (h)} + \frac {1} {12} \; \tilde H_{\mu\nu\eta}^ {(h)} \; 
\tilde H^{\mu\nu\eta (h)} - \frac {\partial}{\partial \theta} \; 
\frac {\partial}{\partial \bar \theta} \;\Bigl [ \frac {i}{2} \; {\cal B}_\mu^{(h)}
\; {\cal B}^{\mu{(h)}} \nonumber\\
&+& \; \frac {1}{2}\; {\cal F}^{(h)} \;\bar {\cal F}^{(h)} 
- \; \frac {1}{4} \; {\cal B}_{\mu\nu} ^{(h)} \; {\cal B}^{\mu\nu (h)}
+ \bar {\cal F}_\mu^{(h)} {\cal F}^{\mu (h)} + 2\; \beta^{(h)} \; \bar \beta^{(h)} \Bigr ], 
\end{eqnarray}
where $\tilde {\cal L}_B$ and $\tilde {\cal L}_{\bar B}$ are the super Lagrangian 
densities defined on the (4, 2)-dimensional supermanifold (without the topological term). 
It is elementary now to note that the above super Lagrangian densities satisfy
\begin{eqnarray}
&& \lim_{\theta \to 0} \; \frac{\partial} {\partial \bar\theta}\; \tilde {\cal L}_B = 0, \qquad
\lim_{\bar \theta \to 0} \; \frac{\partial} {\partial \theta}\; \tilde {\cal L}_B = 0,
\nonumber\\
&& \lim_{\theta \to 0} \; \frac{\partial} {\partial \bar\theta}\; \tilde {\cal L}_{\bar B} 
= 0, \qquad \lim_{\bar \theta \to 0} \; \frac{\partial} {\partial \theta}\; \tilde 
{\cal L}_{\bar B} = 0,
\end{eqnarray}
which capture the (anti-) BRST invariance of the Lagrangian densities (49) and (50)
(without $T (x)$ term) in the physical four dimensions of spacetime.

Now we focus on the (anti-) BRST invariance of the topological $ \bigl[ T (x)  
= (B\wedge F) (x) \bigr]$ term of the 4D Lagrangian densities (49) and (50).
The superfield generalization of this term is given below:
\begin{eqnarray}
T (x) \to \tilde T  (x, \theta, \bar \theta) = \frac {1} {4} \; m \; \varepsilon^{\mu\nu\eta\kappa}
 \; {\cal B}_{\mu\nu}^{(h)} 
(x, \theta, \bar \theta) \; \tilde {\cal F}_{\eta\kappa}^{(h)} (x, \theta, \bar\theta).
\end{eqnarray}
It is clear from the HC for the 1-form gauge theory (cf. (21)) that $\tilde F_{\mu\nu}^{(h)}
(x, \theta, \bar\theta) = F_{\mu\nu} (x)$. Thus, the above topological term can be 
expressed in terms of superfields (after the application of HC) as\footnote{Note that 
we have taken here the positive signs in the expansion of ${\cal B}_{\mu\nu} ^{(h)} 
(x, \theta,\bar \theta)$ to be consistent with our transformations in (38) and (43).}
\begin{eqnarray}
\tilde T (x, \theta, \bar\theta) &=& \frac{1}{4} \; m \; \varepsilon^{\mu\nu\eta\kappa}
\Bigl [B_{\mu\nu} (x) + \theta\; (\partial_\mu \bar C_\nu - \partial_\nu \bar C_\mu)
+ \bar \theta\; (\partial_\mu  C_\nu - \partial_\nu  C_\mu) \nonumber\\
&+& \;\theta \;\bar\theta \;(\partial_\mu B_\nu - \partial_\nu B_\mu) \Bigr ]\; 
F_{\eta\kappa} (x).
\end{eqnarray}
The BRST, anti-BRST and combined (anti-) BRST invariance of the above term can be expressed, 
in the language of the superfield formalism, as
\begin{eqnarray}
&& \lim_{\theta \to 0}\; \frac{\partial}{\partial \bar \theta} \; \tilde T  
(x, \theta, \bar\theta) =
\partial_\mu \;\bigl [ m\; \varepsilon^{\mu\nu\eta\kappa} C_\nu \;\partial_\eta A_\kappa \bigr ] 
\equiv s_b \; (T (x)), \nonumber\\
&& \lim_{\bar \theta \to 0} \;\frac{\partial}{\partial  \theta} \; \tilde T  
(x, \theta, \bar\theta) =
\partial_\mu\; \bigl [ m \;\varepsilon^{\mu\nu\eta\kappa} \bar C_\nu \;\partial_\eta A_\kappa \bigr ]
\equiv s_{ab} \; (T (x)), \nonumber\\
&& \frac{\partial} {\partial \bar \theta}\; \frac{\partial}{\partial \theta} \; \tilde T 
(x, \theta, \bar\theta) = \partial_\mu \;\bigl [ m \;\varepsilon^{\mu\nu\eta\kappa} 
B_\nu \;\partial_\eta A_\kappa \bigr ] \equiv s_b \; s_{ab} \; (T (x)) .
\end{eqnarray}
The above equations imply that the topological term transforms to the total spacetime
derivatives under BRST, anti-BRST and combined (anti-) BRST symmetry transformations.
As a consequence, the action remains invariant under the nilpotent (anti-) BRST symmetry 
transformations.

It is worthwhile to point out that the topological term is somewhat different from the rest 
of the terms of the Lagrangian densities (36) and (37) because it always transforms to a total
spacetime derivative under the gauge and (anti-) BRST symmetry transformations. This is what
is reflected in (61) within the framework of geometrical superfield formalism. It is clear
from equation (58) that the super Lagrangian densities (without the topological term) are
such that their translations along the Grassmannian directions lead to zero result. The
super-topological term (60), however, behaves in a distinct manner because its translation
along the Grassmannian directions (i.e. $\theta, \bar\theta$ and $\theta\bar\theta$) lead
always to a total spacetime derivative term (cf. (61)).\\

\noindent
{\bf 7. Conclusions}\\

\noindent
We have performed the BRST quantization of the 4D topological massive Abelian $U(1)$ gauge model
(in the presence of the celebrated $B \wedge F$ term). Our (anti-) BRST symmetry transformations
(38) and (43) respect a couple of basic requirements  of the BRST formalism because they 
satisfy

(i) the off-shell nilpotency of order two ($s_{(a)b}^2 = 0$), and

(ii) the absolute anticommutativity property ($s_b s_{ab} + s_{ab} s_b = 0$) on the constrained
     surface defined by the field equation (30).

It is the superfield formalism, proposed in [30-39], that has been able 
to help us in achieving the above type of (anti-) BRST symmetry transformations that 
obey the basic requirements of the BRST formalism.

One of the key results of our present investigation is the derivation of the CF-type restriction 
(in the context of the topologically massive Abelian $U(1)$ gauge theory) which enables us to
obtain the absolute anticommutativity of the (anti-) BRST symmetry transformations. It should be
recalled that, for the first time, the CF condition [25] appeared in the BRST description of
the non-Abelian 1-form gauge theory. In our earlier works [40,41], a deep connection between the
CF-type restrictions and the geometrical objects, called gerbes, has been
established. In fact, the existence of the CF-type restriction is 
an inevitable consequence when we exploit the superfield formalism of [30,31] in 
the context of higher $p$-form ($p \ge 2$) gauge theories.

The distinguishing feature of the topological term (i.e. $T = B\wedge F$) becomes quite
transparent in the framework of superfield formalism. In this connection, mention 
should be made that the operation of the Grassmannian derivatives on 
the super-topological expression (59) always yields a total spacetime derivative term.
These expressions, in turn, imply the (anti-) BRST invariance of the topological term
in 4D. This is not the case, however, with the rest of the terms of the Lagrangian density in 4D
(or its counterpart in (4, 2)-dimensional supermanifold) where the (anti-) BRST invariance
ensues because of the (anti-) BRST transformations on all the terms.

The central objective of our present investigation has been to take a modest step in the 
direction to obtain the off-shell nilpotent and absolutely anticommuting (anti-) BRST 
symmetry transformations for the topologically massive version of the 4D {\it non-Abelian} 
gauge theory. Many attempts [8,28,29], in this direction, have already been made. The existence  
of the CF-type condition and the coupled Lagrangian densities have not been deduced,
however, in the above attempts. Our future endeavor [42] would be to obtain the above 
mentioned decisive features (in the context of non-Abelian version of our present model) by
exploiting the superfield formalism proposed by Bonora, etal. [30,31] to obtain
the proper (anti-) BRST symmetries.\\

\noindent
{\bf Acknowledgements}\\

\noindent
Financial support from the Department of Science and Technology, Government
of India, under the SERC project grant No: SR/S2/HEP-23/2006, is gratefully 
acknowledged.

\end{document}